# MultiNoC: A Multiprocessing System Enabled by a Network on Chip

Aline Mello, Leandro Möller, Ney Calazans, Fernando Moraes
*Faculdade de Informática – PUCRS -Av. Ipiranga, 6681, Porto Alegre, Brazil*
{mello, moller, calazans, moraes}@inf.pucrs.br

## Abstract

*The MultiNoC system implements a programmable on-chip multiprocessing platform built on top of an efficient, low area overhead intra-chip interconnection scheme. The employed interconnection structure is a Network on Chip, or NoC. NoCs are emerging as a viable alternative to increasing demands on interconnection architectures, due to the following characteristics: (i) energy efficiency and reliability; (ii) scalability of bandwidth, when compared to traditional bus architectures; (iii) reusability; (iv) distributed routing decisions. An external host computer feeds MultiNoC with application instructions and data. After this initialization procedure, MultiNoC executes some algorithm. After finishing execution of the algorithm, output data can be read back by the host. Sequential or parallel algorithms conveniently adapted to the MultiNoC structure can be executed. The main motivation to propose this design is to enable the investigation of current trends to increase the number of embedded processors in SoCs, leading to the concept of "sea of processors" systems.*

## 1. Design Overview

This work presents the implementation of multiprocessing systems, connected through a NoC. According to ITRS estimation, in 2012, SoCs will have hundreds of IP cores, operating at clock frequencies near 10 GHz. In this context, a Network-on-Chip (NoC) [1] appears as a possible solution for future on-chip interconnections. A NoC is an on-chip network composed by IP cores connected to routers, which are interconnected by communication channels. Another motivation to present this design is the current trend to increase the number of embedded processors in SoCs, leading to the concept of "sea of processors" systems [6].

*MultiNoC* comprises four IP cores connected to the NoC, as illustrated in Figure 1:

- 2 **R8** embedded processors. R8 is a load-store 16-bit processor architecture, containing a 16x16 bit register file, and supporting execution of 36 distinct instructions. Each R8 processor has an attached local memory for program and data (1K 16-bit words), acting as a unified cache.

- 1 **memory** IP, implemented with 4 BlockRAMs, resulting in a capacity of 1024 16-bit words.

- 1 **RS-232** serial IP, providing bi-directional communication with a host computer.

The *MultiNoC* system is a NUMA (non-uniform memory access) architecture, in which each processor has its own local memory, but can also have access to memory owned by other processors, or to remote memory.

The external interface of the *MultiNoC* system comprises 4 signals: (*i*) **reset**, responsible to initialize the *MultiNoC* system; (*ii*) **clock**, a basic synchronization signal; (*iii*) **tx**, data from the host computer to the *MultiNoC* system; (*iv*) **rx**, data from the *MultiNoC* system to the host computer.

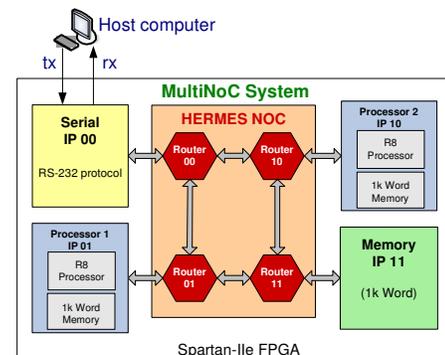

**Figure 1: MultiNoC system block diagram.**

The limitations imposed to the current version of the system arise from the employed FPGA area restrictions, as well as from the choice of using serial low cost, low performance external communication. The approach can be extended to any number of processor IPs and/or memory IPs, using the natural scalability of NoCs. It can also be adapted to faster external interface protocols, such as USB, PCI, Firewire, etc.

## 2. Detailed description

The next Sections present in detail each of the *MultiNoC* IP cores.

## 2.1. Hermes IP core

The Hermes NoC employs *packet switching*, a communication mechanism in which packets are individually routed between cores, with no previously established communication path [5]. The *wormhole* packet switching mode is used to avoid the need for large buffer spaces [9]. The *routing algorithm* defines the path taken by a packet between the source and the destination. The deterministic XY routing algorithm is employed.

The Hermes NoC follows a mesh topology, justified to facilitate routing, IP cores placement and chip layout generation. The routers in *MultiNoC* use an 8-bit flit size, and the number of flits in a packet is fixed at $2^{(\text{flit size in bits})}$. The first and the second flits of a packet are header information, being respectively the address of the target router, named *header flit*, and the number of flits in the packet payload. An asynchronous handshake protocol is used between neighbor routers. The physical interface between routers is composed by the following signals:

- *tx*: control signal indicating data availability;
- *data_out*: data to be sent;
- *ack_tx*: control signal indicating successful data reception.
- *rx*: control signal indicating data availability;
- *data_in*: data to be received;
- *ack_rx*: control signal indicating successful data reception.

The router, shown in Figure 2, is the main component of a NoC, responsible for providing transfer of packets between IPs [10]. The router has a single, centralized control logic and up to five bi-directional ports: East, West, North, South and Local. Each port has an input buffer for temporary storage of packets. The Local port establishes the communication between the router and its local IP. The other ports connect the routers to their neighbor routers.

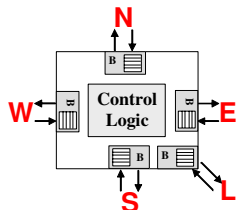

**Figure 2: Hermes router architecture. B indicates input buffers.**

The control logic implements the *routing* and *arbitration* algorithms. When a router receives a header flit, arbitration is performed, and if the incoming packet request is granted, an XY routing algorithm is executed to connect the input port data to the correct output port. If the chosen port is busy, the header flit, as well as all subsequent flits of this packet, will be blocked in the input buffers. The routing request for this packet will remain active until a connection is established in some future execution of the procedure in this router. When the XY routing algorithm finds a free output port to use, the connection between the input port and the output port is established. After routing all flits of the packet, connection is closed. At the operating frequency of 50MHz, with a word size (*flit*) of 8 bits the theoretical peak throughput of each Hermes router is 1Gbits/s.

A router can establish up to five connections simultaneously. Arbitration logic is used to grant access to an output port when one or more input ports require a connection at the same time. A round-robin arbitration scheme is used to avoid starvation.

When a flit is blocked in a given router, the performance of the network is affected, since several flits belonging to the same packet may be blocked in several intermediate routers. To lessen the performance loss, a 2-flit buffer is added to each input router port, reducing the number of routers affected by the blocked flits. Larger buffers can provide enhanced NoC performance. *MultiNoC* employs small buffers to cope with FPGA area restrictions. The inserted buffers work as circular FIFOs.

The minimal latency in clock cycles to transfer a packet from source to destination is given by:

$$latency = \left(\sum\nolimits_{i=1}^{n} R_i\right) + P \times 2$$

where: *n* is the number of routers in the communication path (source and target included), $R_i$ is the time required by the routing algorithm at each router (at least 7 clock cycles), and *P* is the packet size. This number is multiplied by 2 because each flit requires at least 2 clock cycles to be sent, due to the handshake protocol.

The Hermes NoC in the *MultiNoC* system internally supports nine distinct packet formats, which define a set of services offered by the communication network to the IP Cores connected to it. The packets denominations/functions are:

1. *read from memory*, is used to request data from memory;
2. *read return*, response for a read request from memory;
3. *write in memory*, is used to store data into some memory of the system;
4. *activate processor*, initiates the processor, that then starts executing instructions from the first position of its local memory;
5. *printf*, is used by processors to send data to the host computer;
6. *scanf*, is used by processors to request user input data from the host computer;
7. *scanf return*, receives the requested input data from the host computer;
8. *notify*, is used to wake up a processor that has been blocked by a *wait* command;
9. *wait*, blocks a processor until being notified.

## 2.2. Serial IP core

The Serial IP Core is responsible to provide communication between the user working in a host computer and the modules of the system connected through the NoC. This communication is performed by an RS-232 protocol standard serial interface.

Figure 3 presents the Serial IP external interface. The signals at the top of the figure connect the module with the host computer. The signals at the bottom connect the module with the NoC. The function of each signal is:

- *rxd*: receives data serially from the host computer.
- *txd*: sends data serially to the host computer.
- *tx, data_out, ack_tx, rx , data_in, ack_rx*: interface with the HERMES NoC.

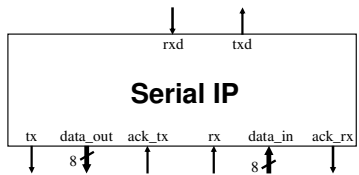

**Figure 3: Serial IP external interface.**

The basic function of the Serial IP is to assemble and disassemble packets. When information comes from the host computer, the Serial IP creates a valid NoC packet. When a packet is received from the NoC it must be disassembled, and sent serially to the host computer.

The Serial IP accepts seven commands. Four commands are handled by the host computer: (*i*) read from memory; (*ii*) write to memory; (*iii*) activate processor; (*iv*) *scanf* return. The other three commands accepted by the serial IP come from the HERMES NoC to the host computer: (*i*) *printf*; (*ii*) *scanf*; (*iii*) *read return*.

## 2.3. Memory IP core

The Memory IP core provides storage for data and/or instructions, and can be accessed through the processor-memory bus or through the NoC. Three Memory IP cores are used in the *MultiNoC* system: two are internally connected to an R8 processor to form the processor IP core and one is an independently accessible remote memory.

Each Memory IP contains 4 BlockRAM modules, each organized as 1024 4-bit words, and control logic to arbitrate the access to the memory banks. Figure 4 shows the external interface of the Memory IP core and the BlockRAMs organization. The access to the memory banks is done in parallel, reading and writing 16-bit words. This access may be done by the processor interface or the NoC interface. The processor interface is not existent in the remote memory IP core.

The memory IP external interface is composed by the following signals:

- *clock*: system clock (not shown in Figure 4).
- *reset*: when asserted, initializes the control logic (not shown in Figure 4).
- *addressCore*: system address provided to this IP (not shown in Figure 4).
- *tx, data_out, ack_tx, rx , data_in, ack_rx*: interface with the HERMES NoC
- interface with the processor:
  - *ceR8*: enables the memory to read/write operations.
  - *rwR8*: read or write operation selection.
  - *addrR8*: processor address bus (16-bit).
  - *dinR8*: processor input data bus (16-bit).
  - *doutR8*: processor output data bus (16-bit).
  - *busyNoCR8*: signals to the memory banks that an operation of the processor with the Hermes NoC is under way (e.g. I/O, read/write).
  - *busyNoCMem*: signals to the processor that an operation of the memory with the Hermes NoC is under way (e.g. return read).

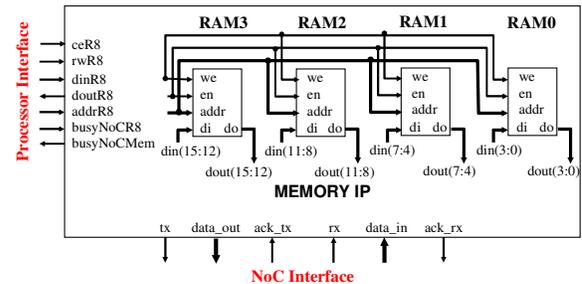

**Figure 4: Memory IP block diagram.**

The *busyNoCR8* and *busyNoCMem* are responsible to prevent the processor and the memory from using the same interface with the NoC simultaneously. The highest priority to access the memory banks is given to the processor.

## 2.4. Processor IP core

The Processor IP external interface, as well as its two main internal modules, is presented in Figure 5. The Processor IP includes the R8 soft core processor [2]; a Memory IP, acting as a unified cache; control logic responsible for interfacing these modules to the HERMES NoC. The R8 soft core processor was chosen due to its simplicity, low area footprint and flexibility to be modified.

The R8 processor is a 16-bit Von Neumann architecture (unified instruction/data memory), with a CPI (Clocks Per Instruction) between 2 and 4.. The datapath contains 16 general purpose registers, an instruction register (IR), program counter (PC), stack pointer (SP), and 4 status flags (negative, zero, carry and overflow).

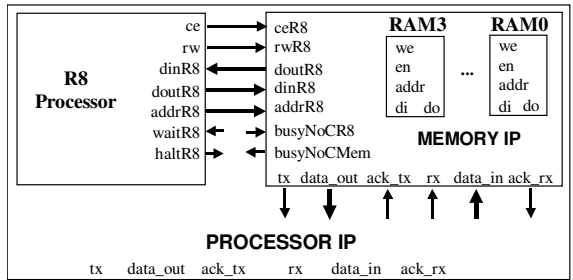

**Figure 5: Processor IP block diagram.**

The Processor IP control logic commands the execution of the R8 processor, putting it in *wait state* each time the processor executes a load-store instruction (see waitR8 signal in Figure 5). Load-store operations can access: (*i*) the local memory; (*ii*) a remote memory; (*iii*) I/O devices; (*iv*) other processors, for synchronization purposes. The next 3 Sections detail these different access modes.

**Memory Accesses**

To determine which device of the *MultiNoC* system the R8 processor is accessing by load-store instructions, address ranges were defined to each memory. The address ranges for the *MultiNoC* system are presented in Figure 6.

```
01 if(address>=0 && address<1024 ){
02         globalAddress = address;
03         addressCore=1; //local memory
04 }
05 else if(address>=1024 && address<2048){
06         globalAddress = 1024 – address;
07         addressCore=2; //other processor
08 }
09 else if(address>=2048 && address<3072){
10         globalAddress = 2048 – address;
11         addressCore=3; //remote memory
12 }
```

**Figure 6: C code illustrating the access to local/remote memories. The Processor IP control logic arbitrates which memory block is accessed.**

**I/O Operations**

The I/O operations are mapped to the FFFFh memory address. Thus, when the ST instruction is executed the *printf* is performed and when the LD instruction is executed the *scanf* is performed.

**Synchronization Operations**

Multiprocessor systems require synchronization mechanisms among processors to implement distributed applications. The synchronization among processors can be done through shared memory or explicit message exchange. The second mechanism was chosen due to the use of NoCs,.

The Processor IP control logic implements the *wait* and *notify* commands, both memory-mapped. The wait command is responsible to block the execution of the processor until the reception of a *notify* command. The *wait* command is identified through the execution of a store instruction (ST) at address FFFEH with the number of the processor that will restart the processor executing ST with a *notify* command. The *notify* command is identified through the execution of ST at address FFFDH with the number of the processor that will be restarted.

For example, when the R8 processor with address *1* executes the instruction *"ST R3, R1, R2"* (R3=2, R1=0, R2=FFFEH), the Processor IP control logic pauses the R8 processor until receiving a packet with a notify command from the IP with address 2. The R8 processor with address 2 should execute *"ST R3, R1, R2"* (R3=1, R1=0, R2=FFFDH) to create the notify command.

## 3. System prototyping

The target device is a Spartan-IIe XC2S200E [11]. The *MultiNoC* system uses 98% of the available slices and 78% of the LUTs. It is important to stress the value of floorplanning in designs using most of the FPGA surface. This generates a complex optimization problem that had to be solved. The use of synthesis and implementation options alone was not sufficient to make the design fit in the restricted area of the XC2S200E device. Even multiple choices of alternate synthesis parameters could not handle the 98% occupation of the design adequately. Figure 7 illustrates the *MultiNoC* system floorplan design that enabled physical synthesis to be successful.

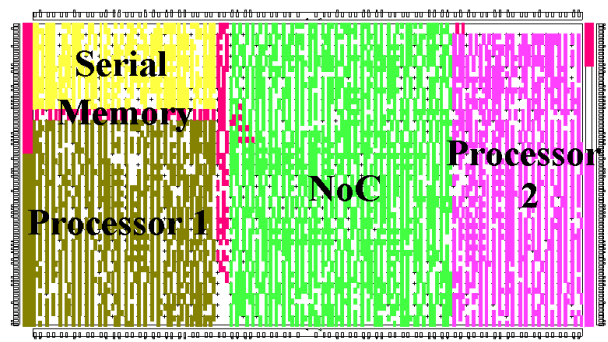

**Figure 7: MultiNoC design floorplan.**

The reasoning behind the placement design of the IPs is justified as follows:
- the NoC IP is placed in the middle of the FPGA, offering easy access to it from all IPs in the system;
- the Serial IP is placed next to the I/O pins responsible for the data transmission/reception to reduce global wire length and routing congestion;
- the Processor IPs are placed in the left/right side of the FPGA, near to the corresponding BlockRAMs for the same reason of the previous item;
- the Memory IP (the smallest IP) is placed in the remaining area.

The original clock of the prototyping board, 50MHz, was divided by two, using a *clkdll* component. The frequency was reduced, due to the delay estimated by the timing analysis tool, 21.23 MHz. Despite the fact that the employed frequency is higher (25 MHz), the circuit worked correctly.

The NoC area can be seen to be an important part of the design when compared to the other IPs. In fact, NoCs trade increased bandwidth (and thus performance) for increased area. However, NoCs are in principle designed for much bigger systems than this prototype. It is not uncommon to consider that NoCs are a feasible communication medium for systems containing more than a hundred IPs (e.g. 10x10 NoCs) [8]. When more area is available, the IPs connected to the NoC can increase in area and functionality. The router surface will remain constant and the NoC dimensions will scale less than the IPs, becoming a very small fraction of the whole system, typically less than 10 or 5%.

## 4. System execution

Figure 8 illustrates the data flow to execute the *MultiNoC* system. The main steps of this flow are detailed below.

- *Simulate the Assembly Code*. The R8 Simulator environment [3] allows writing, simulating and debugging assembly code, generating automatically the object code that must be opened in the current version of the *Serial* software to send it to the R8 processor. Unfortunately, the R8 Simulator is not able to simulate a multiprocessed application.
- *Start the Serial Software*. The *Serial* software [4] enables the host computer to communicate with the Spartan-IIe device.
- *Synchronize SW/HW*. The *MultiNoc* system must receive from the *Serial* software the host computer baud rate, to correctly receive/send data. This is achieved transmitting the value 55H to the *MultiNoc* system.
- *Send Generated Object Code*. The text file obtained after the application simulation is sent to the *MultiNoc* system using the *Serial* software.
- *Fill Memory Contents*. Optionally, data may be also sent to the remote memory or the local memories of the *MultiNoC*. This process of sending data to the memory is similar to sending the object code to the instruction memory.
- *Activate Processors*. The system is now ready to execute the application, since program and data are stored. This action is achieved by sending a specific command through the *Serial* software (activate processor command).

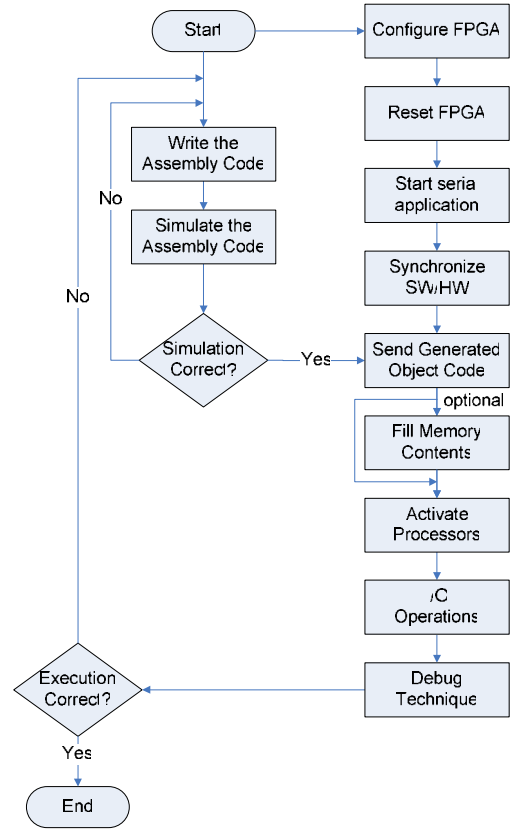

**Figure 8: *MultiNoC* system flow diagram.**

- *I/O Operations*. In order to execute I/O operations, the *Serial* software has interaction monitors for each processor. Figure 9(2) presents a monitor with input/output operations indicating the result of an application execution.

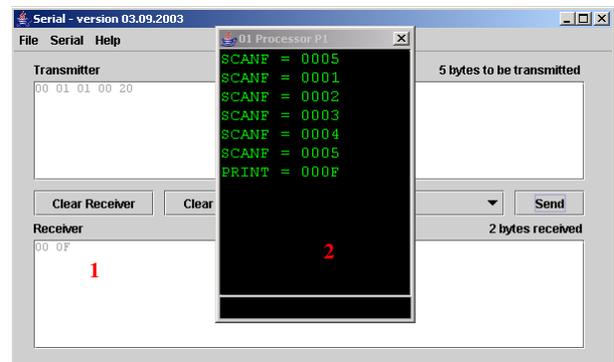

**Figure 9: Two ways of debugging the *MultiNoC* system: 1) read operation; 2) *printf* operation.**

- *Debug*. There are two ways of verifying the prototype correct functionality. One is directly reading memory values (step 1 in Figure 9) and the other is using *printf* instructions executed by some processor (step 2 in Figure 9). *Printf* instructions are part of the application code and can help verifying intermediate values.

Memory reads can be used to verify the memory contents at the end of the execution. In the example presented in Figure 9, the user has typed "00 01 01 00 20", meaning a read operation (00) from P1 processor local memory (01), reading just one memory position (01) and starting at address 0020H.

The software of Figure 9 is used for basic communication with *MultiNoC* system. More complex applications have been developed. One example is a parallel edge detection which uses the basic functions of the *Serial* software to construct more powerful user interfaces. For demonstration purposes, consider the GUI presented in Figure 10. In this application the host computer sends an image line, after what each embedded processor computes one gradient (gx and gy). Next, that embedded processor adds gx and gy and notifies the host, which receives the processed line, and sends a new line to the *MultiNoC* system.

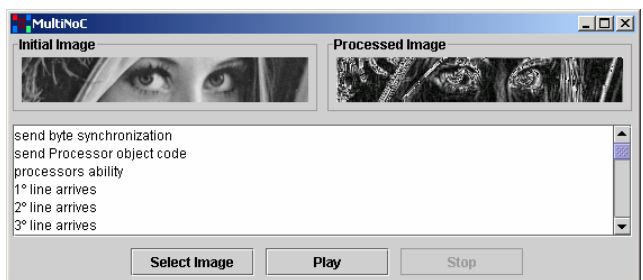

**Figure 10: Parallel edge detection GUI.**

## 5. Conclusions and future work

The *MultiNoC* system is in fact an exercise of implementing and making available a design platform on top of which applications can be effectively and rapidly prototyped. This is indeed a recently proposed new design paradigm called *platform based design*, as opposed to existing paradigms, *system level design* and *component based design* [7].

Future research with the MultiNoC system includes the development of a multiprocessor simulator. This tool is important to detect distributed application errors and to synchronize software running on different processors. Another important tool is a C compiler to automatically generate R8 assembly code, allowing faster software implementation.

Mapping the *MultiNoC* system in a larger FPGA device would allow increasing the NoC dimension and the number of IPs connected to it. This new system can be composed by more instances of the presented pre-designed and pre-verified IP cores, adopting the concept of design reuse, or by implementing new IP cores. Increasing the number of identical IPs enhances the parallelism degree. On the other hand, increasing the amount of different IPs contributes with new functionalities to the *MultiNoC* system.

One of the current research foci is on partial and dynamic reconfiguration applied to the *MultiNoC* system. Partial and dynamic reconfiguration allows, for example, that the IP cores position be modified in execution at runtime, favoring the IPs communication with improved throughput. Reconfiguration can also be used to reduce system area consumption through insertion and removal of IP cores on demand.

The *MultiNoC* system can be very useful to undergraduate/graduate students for learning concepts arising in hardware description languages, distributed systems, parallel processing hardware development environments, prototyping designs and applying digital systems concepts to related disciplines. Additionally, the *MultiNoC* system has a low area overhead, being able to be prototyped in small devices, decreasing the cost to be acquired in large numbers for academic laboratories.

## 6. Acknowledgements

This research has been partially funded by CNPq through grants 550009/2003-5 and 307665/2003-2.